\documentclass[reprint,prl,letterpaper,superscriptaddress]{revtex4-1}
\usepackage[T1]{fontenc} 
\usepackage[applemac]{inputenc}
\usepackage{graphicx}
\usepackage{dcolumn}
\usepackage{bm}
\usepackage{amsmath,amssymb}
\usepackage{pifont}
\usepackage{graphicx}

\usepackage{epstopdf}
\usepackage{bm}
\usepackage[normalem]{ulem} 
\usepackage{gensymb} 
\usepackage[usenames,dvipsnames]{color}

\definecolor{OliColor}{rgb}{0.1,0.1,0.8} 
\definecolor{OliComColor}{rgb}{0.5,0.5,1} 
\def\Oc#1{}

\definecolor{AbeColor}{rgb}{0.8,0.1,0.1} 
\definecolor{AbeComColor}{rgb}{1,0.5,0.5} 

\def\OV#1{#1}
\def\OB#1{#1}

\begin{document}

\graphicspath{{Figures_v17/}}

\title{Drying by Cavitation and Poroelastic Relaxations in Porous Media with Macroscopic Pores Connected by Nanoscale Throats}

\author{Olivier Vincent}
\email{orv3@cornell.edu}
\affiliation{School of Chemical and Biomolecular Engineering, Cornell University, Ithaca, New York 14853, USA}

\author{David A. Sessoms}
\affiliation{School of Chemical and Biomolecular Engineering, Cornell University, Ithaca, New York 14853, USA}

\author{Erik J. Huber}
\affiliation{Sibley School of Mechanical and Aerospace Engineering, Cornell University, Ithaca, New York 14853, USA}

\author{Jules Guioth}
\affiliation{School of Chemical and Biomolecular Engineering, Cornell University, Ithaca, New York 14853, USA}

\author{Abraham D. Stroock}
\email{ads10@cornell.edu}
\affiliation{School of Chemical and Biomolecular Engineering, Cornell University, Ithaca, New York 14853, USA}
\affiliation{Kavli Institute at Cornell for Nanoscale Science, Cornell University, Ithaca, New York 14853, USA}


\begin{abstract}
We investigate the drying dynamics of porous media with two pore diameters separated by several orders of magnitude. Nanometer-sized pores at the edge of our samples prevent air entry, 
while drying proceeds by heterogeneous nucleation of vapor bubbles – cavitation – in the liquid in micrometer-sized voids within the sample. 
We show that the dynamics of cavitation and drying are set by the interplay of the deterministic poroelastic mass transport in the porous medium and the stochastic nucleation process. Spatio-temporal patterns emerge in this unusual reaction-diffusion system, with temporal oscillations in the drying rate and variable roughness of the drying front. 
\end{abstract}

\maketitle



\begin{figure}[]
	\begin{center}
	\includegraphics{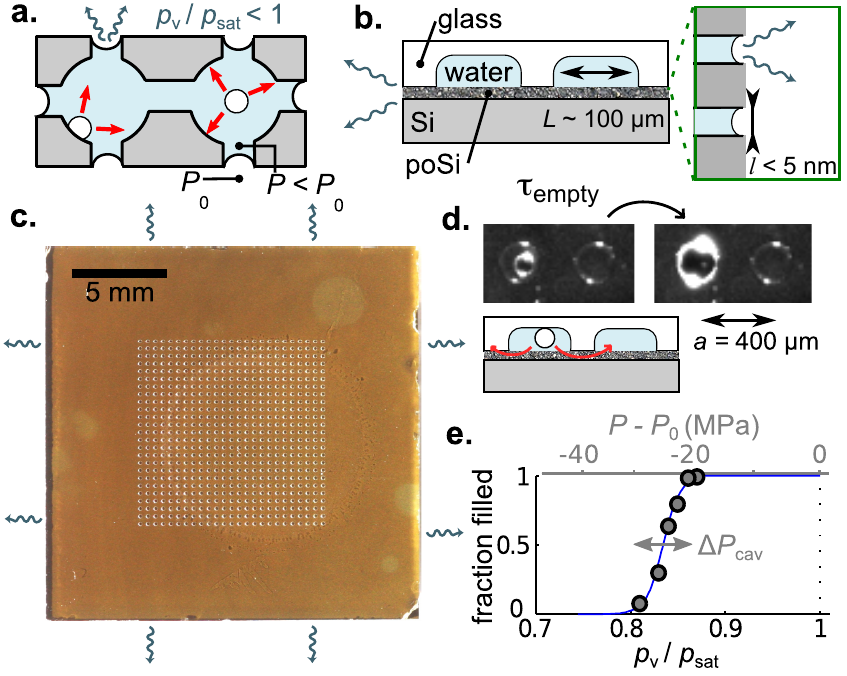}
			\caption{\small   (a) A porous medium with ink-bottle morphology may dry from inside by cavitation. (b) Schematic side-view of our extreme ink-bottle geometry with microfabricated voids interconnected by porous silicon (poSi). (c) Photograph showing top-view of a sample with an array of $25\times25$ voids (bright dots) bonded to a layer of nanoporous silicon (orange background). 
Each void is dome-shaped with diameter of $200 \, \mathrm{\mu m}$ and depth of $27  \, \mathrm{\mu m}$.
(d) Optical micrographs of a cavitation event (top view) and schematic drawing of the water fluxes (side view). Left micrograph shows the bubble immediately after cavitation; right micrograph shows the empty void. (e) Survival curve (see text). 
\label{fig:Fig1}
}
	\end{center}
\end{figure}

The desorption, or drying, of liquids from porous media is important in a variety of contexts, both in nature (e.g., the movement of water in plants \cite{Stroock2013}) and in technology (e.g., the synthesis \cite{SolGelScience} and characterization \cite{BJH_method_1951} of advanced materials). Studies of desorption have focused on the dynamics and pattern formation associated with drying \cite{Shaw1987,Prat2002,LehmannOr2008} and the thermodynamics that define the shapes of desorption isotherms \cite{Wallacher2004,Ravikovitch2002}. Despite this attention, uncertainties remain regarding the physical processes that govern desorption.
It has been proposed  that, rather than by
receding of the liquid phase from the edges of the material, drying from porous media could occur by cavitation, i.e. 
\OB{the spontaneous formation of vapor bubbles 
either when the liquid tensile strength is attained  \cite{Schofield1948} or by thermally activated nucleation in the metastable pore liquid  (Fig. \ref{fig:Fig1}a) \cite{Or2002}.}
This process has been observed in simulations \cite{Sarkisov2001,Ravikovitch2002} and has been proposed on several occasions to explain the shape of desorption isotherms in nano-scale porous media \cite{Ravikovitch2002,Grosman2011,Bonnet2013}, or the apparent emergence of drying events far from the evaporation front \cite{Sarkar1994,Scherer1995}. Yet, we are unaware of direct \OB{optical} observation of \OB{desorption} by cavitation or of an investigation of its effect on \OB{drying dynamics}.


In this Letter, we present a tailored porous medium formed of microfabricated voids coupled to each other and the outside via a nanoporous substrate (Figs. \ref{fig:Fig1}b-\ref{fig:Fig1}c). This extreme ink-bottle structure -- large pore bodies connected via narrow throats \cite{Cohan1944} -- has allowed us to observe the nucleation and growth of cavitation bubbles during drying (Fig. \ref{fig:Fig1}d) and to show that this process gives rise to interesting \OV{coupled} \OB{drying dynamics} \OV{that is tunable with geometry}.

Drying occurs when a saturated porous medium is placed in a sub-saturated atmosphere (relative humidity $p_\mathrm{v}/p_\mathrm{sat}<1$, where $p_\mathrm{v}$ and $p_\mathrm{sat}$ are the vapor pressure and its saturation value). At high relative humidity, evaporation results in the formation of menisci in the pores at the surface of the material, until local mechanical and chemical equilibrium are established, which occurs when the liquid pressure $P$ satisfies \cite{Wheeler2008}:
\begin{equation}
P-P_0=\frac{RT}{v_\mathrm{m}}\ln\left(\frac{p_\mathrm{v}}{p_\mathrm{sat}}\right)
\label{eq:Kelvin}
\end{equation}
and 
\begin{equation}
P-P_0> -\frac{2\sigma\cos\theta_\mathrm{r}}{r_\mathrm{p}},
\label{eq:Laplace}
\end{equation}
where $v_\mathrm{m} = 1.805\times10^{-5}\, \mathrm{m^3/mol}$ is the molar volume of the liquid, $RT$ is the thermal energy, $\sigma$ is the liquid surface tension, and $\theta_\mathrm{r}$ is the receding contact angle of the mensicus in pores of radius $r_\mathrm{p}$.  At lower relative humidities, for which the pressure difference defined by (\ref{eq:Kelvin}) violates the inequality 
 (\ref{eq:Laplace}), the menisci will recede into the pores.  This scenario is the classical process of drying by capillary invasion from the edges of the material \cite{Scherer1990}.

Drying may also occur by a distinct mechanism: Eq. (\ref{eq:Kelvin}) states that the pressure in the liquid phase is reduced relative to the standard pressure $P_0$, such that it is metastable with respect to the homogeneous or heterogeneous nucleation of bubbles of air and vapor
\cite{Schofield1948,Debenedetti1996} (Fig. 1a).
This mode of drying by cavitation may be favored in “ink-bottle” geometries in which small pores prevent invasion of the menisci and permit the development of large stresses within the pore liquid (“pore-blocking” via Eq. 2), while larger pores inside allow the nucleation of bubbles (Fig. \ref{fig:Fig1}a). 

Here, we report on an extreme version of the ink-bottle geometry in which nanometer- and micrometer-sized voids coexist; previous studies of bi-disperse porous media have been limited to small ($\leq 2$-fold) differences in pore size \cite{Wallacher2004,Bruschi2010,Grosman2011}.
Figs. \ref{fig:Fig1}b and \ref{fig:Fig1}c present our system 
\footnote{See Supplemental Material}: we formed discrete, micrometer-scale voids in the surface of glass via lithography; we formed a $5\,\mathrm{\mu m}$-thick layer of interconnected, nanometer-scale pores ($1\--2\,\mathrm{nm}$ in radius) in the surface of silicon by anodization; and we bonded the glass and silicon anodically to couple the micro-voids to each other and the outside environment via the nano-pores.  

We filled
the evacuated samples by submerging them in liquid water at elevated pressure, then
allowed them to dry in closed chambers with a controlled humidity, and followed the resulting dynamics by time-lapse photography. 
Equations (\ref{eq:Kelvin}) and (\ref{eq:Laplace}) were always verified in our experiments, leaving cavitation as the only plausible mechanism for drying 
\cite{Note1}. As seen in  Fig. \ref{fig:Fig1}d, we could follow discrete cavitation events within the micro-voids during drying.

We first characterized the stability of the bulk liquid water within the micro-voids by lowering the vapor pressure in step increments and counting the number of full voids after $1$ day of equilibration (Fig. \ref{fig:Fig1}e). Across samples, the shape of the survival curves were similar, with 
probability $1/2$ of cavitation for $p_\mathrm{v}/p_\mathrm{sat}=0.83\--0.86$,
corresponding to liquid pressures in the range $P_\mathrm{cav}=-20$ to $-24$ MPa from Eq. (\ref{eq:Kelvin}) (width $\Delta P_\mathrm{cav}=2\--5$ MPa).
This value is less negative than expected for homogeneous nucleation in water ($P_\mathrm{cav}\sim-140$ MPa) \cite{Debenedetti1996}, suggesting a heterogeneous process, but matches closely most measurements of $P_\mathrm{cav}$ in water \cite{Caupin2006}. The exact nucleation mechanism is however not important for the dynamics of drying.


\begin{figure}[]
	\includegraphics{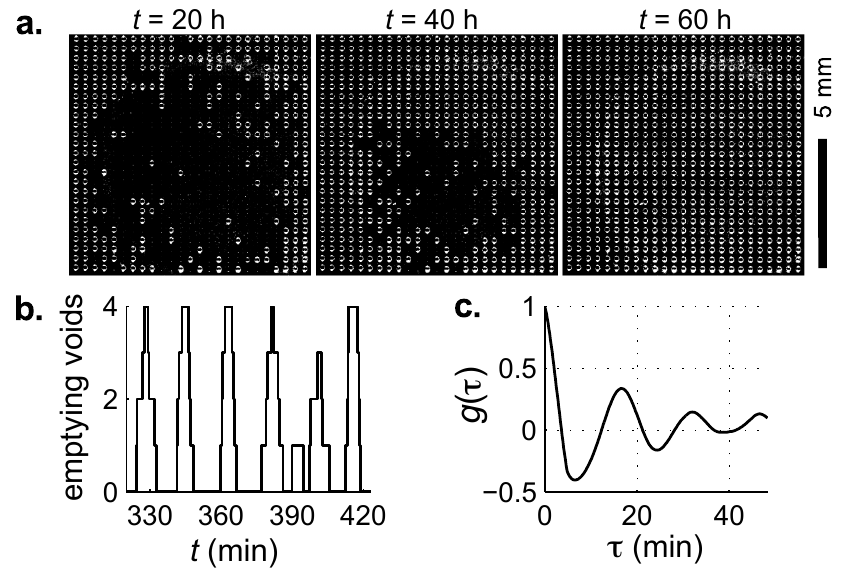}
			\caption{\small Drying dynamics. (a) Optical micrographs showing state of voids in a sample as in Fig. 1c (only array of voids was imaged) during drying at $p_\mathrm{v}/p_\mathrm{sat}=0.84$. Empty voids appear bright. 
Evaporation occurred from all four edges of the samples.
(b) Number of voids in the process of emptying as a function of time. (c) Autocorrelation of the time series as in (b) for all 625 cavitation events.
			\label{fig:Fig2}}
\end{figure}


Fig. \ref{fig:Fig2}a presents a typical drying sequence for $p_\mathrm{v}/p_\mathrm{sat}=0.84$ such that the boundary liquid pressure was $P_\mathrm{b}\simeq-24$ MPa (Eq. \ref{eq:Kelvin}). Drying proceeded in a mixed fashion, with both coherent, front-like progression from the edges to the center and discrete events far away from the front. This observation suggests that the progression of cavitation was controlled by an interplay of deterministic mass transport and stochastic cavitation events that allow drying deep within the sample.

More surprisingly, despite the steady nature of the applied driving force, we found that cavitation and emptying of the voids proceeded in a punctuated manner, with bursts of 1 to $\sim7$ events separated by periods with no events (Fig. \ref{fig:Fig2}b). This progression of bursts \OV{persisted} until all voids were empty \cite{Note1} and showed long time correlations
(Fig. \ref{fig:Fig2}c). This emergent behavior suggests links to the dynamics in other out-of-equilibrium systems \cite{Cross1993} such as the periodic relaxations in a continuously driven sand pile \cite{Jaeger1989}, which motivates the search for mechanisms of positive and negative feedback.

We considered but could exclude positive feedback based on acoustic emissions from cavitation events \cite{Note1}. The period between successive bursts ($\simeq15\,\mathrm{min}$, see
Fig. \ref{fig:Fig2}b-c) was similar to the timescale for bubble growth after cavitation ($\tau_\mathrm{empty}\simeq3.5$ min, see Fig. \ref{fig:Fig1}d and \cite{Note1}), which suggests that the bursting behavior arose from a hydraulic coupling due to the release of water into the surrounding medium during emptying. 


\begin{figure}[]
	\begin{center}
	\includegraphics{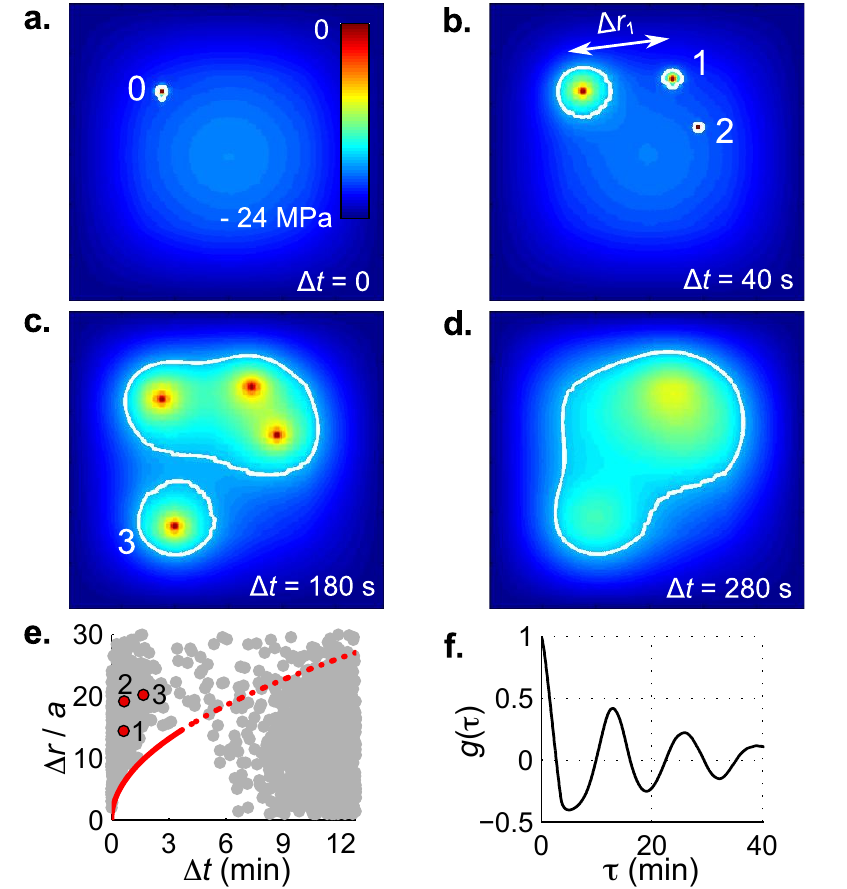}
			\caption{\small  
			Time evolution of the simulated pressure field in a $25\times25$ voids sample as in Fig. \ref{fig:Fig2}a. (a) First cavitation event within a burst, labeled 0. (b) Growing exclusion zones (tracked by white contour lines at $-17$ MPa) (c) Exclusion zones merge. (d) Retraction of the exclusion zones (e) Spatio-temporal representation of the drying dynamics for the full simulation, showing spatial ($\Delta r$) versus temporal ($\Delta t$) separations for all pairs of cavitation events. Events 1-2-3 from (a-d) (with event 0 at origin) are represented as red dots. The red line (continuous for $\Delta t<\tau_\mathrm{empty}$, dashed afterwards) corresponds to $\Delta r = \alpha \sqrt{C\,\Delta t}$ with $\alpha=1.7$. (f) Autocorrelation function of all emptying events in frame (e). 
			\label{fig:Fig3}}
	\end{center}
\end{figure}


In order to check this hypothesis, we developed a model coupling deterministic mass transport and stochastic cavitation kinetics. We treat fluid transport within the porous silicon as poroelastic diffusion \cite{Biot1941}:
\begin{equation}
\frac{\partial P}{\partial t} = C\nabla^2 P
\label{eq:Poroelasticity}
\end{equation}
where $C=\kappa K_\mathrm{liq} / \phi=5\times10^{-8}\,\mathrm{m^2/s}$ is the poroelastic diffusivity, $\kappa = 1.44\times 10^{-17},\mathrm{m^2/(Pa.s)}$ is the Darcy permeability of porous silicon, $\phi=0.6$ is its porosity, and $K_\mathrm{liq}=2.2$ GPa is the bulk modulus of liquid water.

We describe cavitation kinetics in \OV{each void} by classical nucleation theory \cite{Debenedetti1996,Caupin2006}, leading to the following expression \OV{for the rate of nucleation}:
\begin{equation}
\OV{k_\mathrm{cav}(t)}=\Gamma_0 V \exp\left(-\frac{E_\mathrm{nucl}\OV{(P)}}{k_\mathrm{B}T}\right)
\label{eq:CNT}
\end{equation}
where $\Gamma_0$ is a kinetic prefactor, $V$ is the void volume and $k_\mathrm{B}$ is Boltzmann's constant. In the energy barrier for nucleation, $E_\mathrm{nucl}=16\pi\sigma^3 / \left[ 3(P-p_\mathrm{sat})^2 \right]$, $\sigma\simeq0.02$ N/m is an effective surface tension with a value typical of those used to match the observed kinetics of cavitation of water in a variety of experiments \cite{Caupin2006,Wheeler2008,Davitt2010}.  In order to account for spatial heterogeneities, we also allowed $\sigma$ to vary with a narrow distribution ($\Delta\sigma / \sigma\sim0.02\--0.07$). We adjusted $\sigma$ and $\Delta \sigma$ to fit the experimental survival curves (Fig. \ref{fig:Fig1}e, blue line).

We numerically solved Eq. (\ref{eq:Poroelasticity}) with an explicit finite difference scheme \cite{Crank1979} for the evolution of the pressure field, $P(x,y,t)$. The four edges of the sample were maintained at a fixed liquid pressure  $P_\mathrm{b}$ set by the outside humidity (Eq. \ref{eq:Kelvin}).
At each time step $\Delta t$, the probability of nucleation $k_\mathrm{cav} \times \Delta t$ was calculated in each void (Eq. 4), and cavitation was triggered with a Monte Carlo scheme. If cavitation occurred, the pressure in the void was set to $P=p_\mathrm{sat}$ (liquid-vapor equilibrium) until the void was empty of liquid. After emptying, the vapor pressure in the void was allowed to evolve in local  equilibrium (Eq. \ref{eq:Kelvin}) with the liquid pressure $P$ in the neighboring region.

The sequence in Fig. \ref{fig:Fig3}a-d presents snapshots of the calculated pressure field during a burst of four cavitation events under conditions matching the experiment shown on Fig. \ref{fig:Fig2} \footnote{See Supplemental Movie}. When a void cavitates, its pressure jumps from the negative value in the surrounding matrix to $p_\mathrm{sat}$ (red pixels in Fig. \ref{fig:Fig3}a). This transition generates outward flow, emptying of the void, and the growth of a resaturated zone in the surrounding matrix (Fig. \ref{fig:Fig3}b, white lines). Outside of this zone, strong metastability persists and further cavitation events can occur (Fig. \ref{fig:Fig3}b, events 1 and 2); inside this zone, metastability is reduced and the nucleation rate is negligible. The resaturated zones of multiple events grow and merge (Fig. \ref{fig:Fig3}c) until further cavitation is suppressed throughout the sample (Fig. \ref{fig:Fig3}d). Once the cavitated voids have emptied, they cease to act as sources for the rehydration of their surroundings; the pressure in the sample begins to decrease again due to continued evaporation at the boundary, and a new cycle of cavitation (that is, a burst) can start.

Fig. \ref{fig:Fig3}e presents the spatio-temporal statistics of all cavitation events during the simulation. A pair of cavitation events separated by a time $\Delta t$ can only exist outside of an exclusion zone (red line), the spatial extent of which grows as $\Delta r_\mathrm{excl} \sim \sqrt{C\,\Delta t}$ during a timescale on the order of the average emptying time $\tau_\mathrm{empt}$. The growth in $t^{1/2}$ arises from 
poroelastic diffusion (Eq. \ref{eq:Poroelasticity}).

The similarity of the autocorrelation function \OB{(which measures the bursting behavior)} of simulated events (Fig. \ref{fig:Fig3}f) to that observed experimentally (Fig. \ref{fig:Fig2}c) supports the conclusion that the
time-dependent response emerges due to this \emph{suppressive} effect based on hydraulic coupling. As will be shown below, this coupling can be tuned experimentally by varying the geometry of the ink-bottle system.

The emergence of this complex response from the coupling of diffusion (Eq. \ref{eq:Poroelasticity}) and non-linear reaction (Eq. \ref{eq:CNT}) points to a particular analogy with the relaxation of metastable states in reaction-diffusion dynamics that leads, for example, to pattern formation during the precipitation of supersaturated solutions \cite{Stern1954}.


\begin{figure}[]
	\begin{center}
	\includegraphics{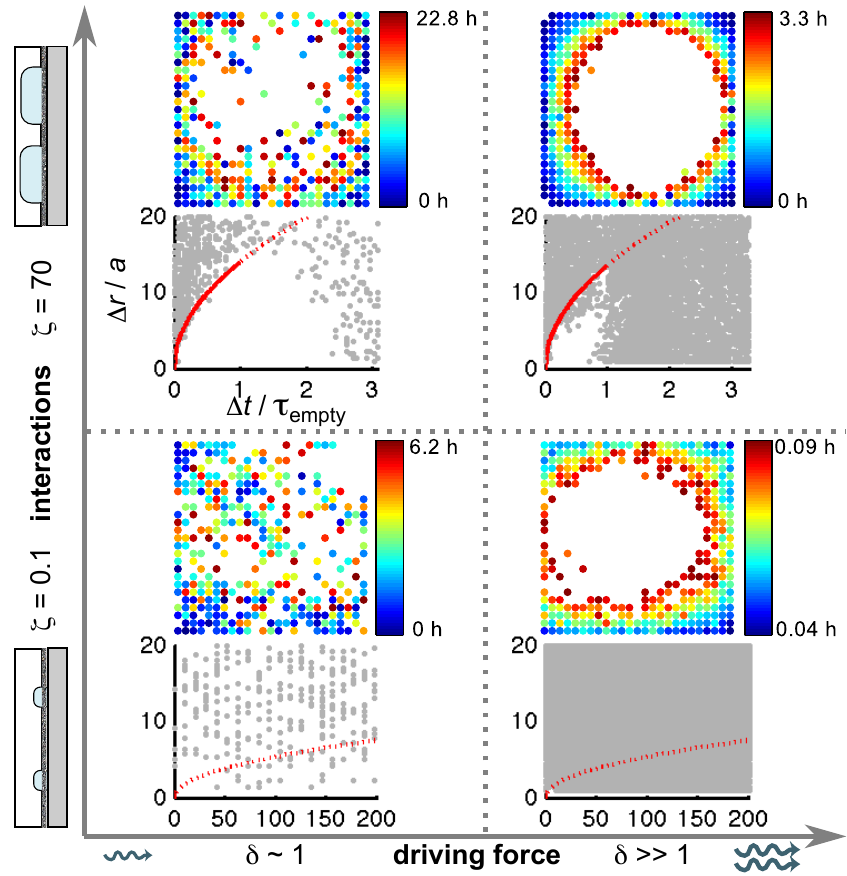}
			\caption{\small
			Experimental state diagram of drying dynamics as a function of driving force and interactions between voids. The upper diagram in each quadrant is a spatial representation of the drying front when half of the voids had cavitated (color indicates event time).
The lower diagram is the measure of temporal correlations introduced in Fig. \ref{fig:Fig3}e (the different datapoint densities reflect the impact of geometry or driving force on drying speed). The top left experiment is the one reported on Fig. 2. 
			\label{fig:Fig4}}
	\end{center}
\end{figure}


We turn to the identification of grouped parameters that control spatio-temporal drying patterns in these ink-bottle materials, with pressure diffusion and cavitation kinetics as the basic ingredients.
First, we define a non-dimensional interaction between voids to capture the negative feedback illustrated in Fig. \ref{fig:Fig3}. 
If the spatial range of the poroelastic coupling $(\Delta r_\mathrm{excl}\sim\sqrt{C\times\tau_\mathrm{empt}})$ is smaller than the distance $a$ between voids, the interaction should vanish.
We thus define a non-dimensional parameter comparing these two lengthscales
\begin{equation}
\zeta = \tau_\mathrm{empt}\times C/a^2,
\label{eq:ND_Interaction}
\end{equation}
which is also equal to the ratio of emptying time to the time $\tau_\mathrm{diff}=a^2/C$ for pressure information to travel from one void to its neighbor.

Second, we define a non-dimensional driving force for cavitation :
\begin{equation}
\delta = (P_\mathrm{cav}-P_\mathrm{b})/\Delta P_\mathrm{cav}
\label{eq:ND_DrivingForce}
\end{equation}
where $P_\mathrm{b}$ is the liquid pressure at the boundary defined by the external humidity 
(Eq. \ref{eq:Kelvin}), and $P_\mathrm{cav}$ and $\Delta P_\mathrm{cav}$ parametrize the survival curve (Fig. 1e). 
We expect $\delta$ to be a natural measure of the roughness of the drying front \cite{Note1}: for strong driving forces ($\delta\gg1$)
we expect sharp fronts, controlled by the diffusion of the boundary pressure into the sample; for weak driving forces (small positive and negative values of $\delta$), we expect spatially random events controlled by the kinetics of cavitation.

We investigated \OV{experimentally} the effect of the \OV{two dimensionless parameters established above}. First, the coupling parameter $\zeta$ (Eq. \ref{eq:ND_Interaction}) \OV{was varied by using} samples with the same inter-void spacing, $a=400\,\mathrm{\mu m}$, but with a void volume that was $\sim 200$-fold smaller than that of the samples discussed up to this point. This 
geometrical change lowered the interaction parameter from $\zeta=70$ (Fig. \ref{fig:Fig4}, top row) to $\zeta=0.1$ (Fig. \ref{fig:Fig4}, bottom row) by decreasing the emptying time by a factor 700. As predicted, the negative feedback between cavitation events was dramatically reduced, as can be seen in the nearly complete disappearance of the exclusion zones for both  the strong and low driving force cases (Fig. \ref{fig:Fig4}, bottom). Further, the time sequence of cavitation showed no measurable temporal correlations. 

Second, to investigate the impact of driving force $\delta$, we exposed the same samples to different humidities (Fig. \ref{fig:Fig4}, $\delta$ increases from left to right).
As predicted, we observed that the drying front became much sharper for higher values of $\delta$, with little effect on the hydraulic coupling (as seen in spatio-temporal maps). This evolution of the front shape is similar to that reported for evaporative drying of porous media \cite{Shaw1987,Prat1999}. The important distinction here is that drying occurs via cavitation at localized, spatially random points beyond the drying front, without percolation of a vapor path to the surface. Our experiments thus provide the first direct, experimental support of the proposal that this evolution of the pattern of drying with the driving force may result from cavitation \cite{Sarkar1994,Scherer1995}.

Finally, we note that our simulations based on Eqs (3) and (4) reproduce the state diagram in Fig. 4 well
\cite{Note1},
providing further support for the relevance of our model.

In conclusion, we have used extreme ink-bottle geometries to provide \OB{direct optical observation} of drying of a porous medium by cavitation. We have further elucidated aspects of the emergent dynamics of this process and shown that we can explain them quantitatively with a model that couples poroelastic transport and cavitation kinetics. This mechanism represents a new route to a non-linear reaction-diffusion process, with mechanical rather than molecular diffusion, and with 
non-linear kinetics coming from nucleation rather than from reaction.
We expect that similar coupled dynamics should occur in highly heterogeneous porous media, such as the ones found in rocks or plant tissues. In the geological context, the propagation of drying stresses could lead to patterns of nucleation of gases from supersaturated solutions that are formed, for example, during the sequestration of 
carbon dioxide \cite{Huppert2014}. In plant tissues, the hydraulic interactions upon drying and cavitation might play a role in the physiological response to drought, as suggested by early results from Dixon et al. \cite{Dixon1984}. We expect that microfabricated structures of the type presented here will provide useful avenues for the exploration of these and other phenomena.

\section*{Acknowledgements}
The authors would like to thank Eugene Choi for the measurement of nitrogen isotherms.
This work was supported by the National Science Foundation	(CBET-0747993 and CHE-0924463), the Air Force Office of Scientific Research (FA9550-09-1-0188), and the Camille Dreyfus Teacher-Scholar Awards program, and was performed in part at the Cornell NanoScale Facility, a member of the National Nanotechnology Infrastructure Network (National Science Foundation; Grant ECCS-0335765).

\bibliography{CoupledCavitationReferences}

\begin{thebibliography}{32}%
\makeatletter
\providecommand \@ifxundefined [1]{%
 \@ifx{#1\undefined}
}%
\providecommand \@ifnum [1]{%
 \ifnum #1\expandafter \@firstoftwo
 \else \expandafter \@secondoftwo
 \fi
}%
\providecommand \@ifx [1]{%
 \ifx #1\expandafter \@firstoftwo
 \else \expandafter \@secondoftwo
 \fi
}%
\providecommand \natexlab [1]{#1}%
\providecommand \enquote  [1]{``#1''}%
\providecommand \bibnamefont  [1]{#1}%
\providecommand \bibfnamefont [1]{#1}%
\providecommand \citenamefont [1]{#1}%
\providecommand \href@noop [0]{\@secondoftwo}%
\providecommand \href [0]{\begingroup \@sanitize@url \@href}%
\providecommand \@href[1]{\@@startlink{#1}\@@href}%
\providecommand \@@href[1]{\endgroup#1\@@endlink}%
\providecommand \@sanitize@url [0]{\catcode `\\12\catcode `\$12\catcode
  `\&12\catcode `\#12\catcode `\^12\catcode `\_12\catcode `\%12\relax}%
\providecommand \@@startlink[1]{}%
\providecommand \@@endlink[0]{}%
\providecommand \url  [0]{\begingroup\@sanitize@url \@url }%
\providecommand \@url [1]{\endgroup\@href {#1}{\urlprefix }}%
\providecommand \urlprefix  [0]{URL }%
\providecommand \Eprint [0]{\href }%
\providecommand \doibase [0]{http://dx.doi.org/}%
\providecommand \selectlanguage [0]{\@gobble}%
\providecommand \bibinfo  [0]{\@secondoftwo}%
\providecommand \bibfield  [0]{\@secondoftwo}%
\providecommand \translation [1]{[#1]}%
\providecommand \BibitemOpen [0]{}%
\providecommand \bibitemStop [0]{}%
\providecommand \bibitemNoStop [0]{.\EOS\space}%
\providecommand \EOS [0]{\spacefactor3000\relax}%
\providecommand \BibitemShut  [1]{\csname bibitem#1\endcsname}%
\let\auto@bib@innerbib\@empty
\bibitem [{\citenamefont {Stroock}\ \emph {et~al.}(2013)\citenamefont
  {Stroock}, \citenamefont {Pagay}, \citenamefont {Zwieniecki},\ and\
  \citenamefont {Holbrook}}]{Stroock2013}%
  \BibitemOpen
  \bibfield  {author} {\bibinfo {author} {\bibfnamefont {A.~D.}\ \bibnamefont
  {Stroock}}, \bibinfo {author} {\bibfnamefont {V.~V.}\ \bibnamefont {Pagay}},
  \bibinfo {author} {\bibfnamefont {M.~A.}\ \bibnamefont {Zwieniecki}}, \ and\
  \bibinfo {author} {\bibfnamefont {N.~M.}\ \bibnamefont {Holbrook}},\ }\href
  {http://dx.doi.org/10.1146/annurev-fluid-010313-141411} {\bibfield  {journal}
  {\bibinfo  {journal} {Annu. Rev. Fluid Mech.}\ }\textbf {\bibinfo {volume}
  {46}} (\bibinfo {year} {2013})}\BibitemShut {NoStop}%
\bibitem [{\citenamefont {Brinker}\ and\ \citenamefont
  {Scherer}(1990)}]{SolGelScience}%
  \BibitemOpen
  \bibfield  {author} {\bibinfo {author} {\bibfnamefont {C.~J.}\ \bibnamefont
  {Brinker}}\ and\ \bibinfo {author} {\bibfnamefont {G.~W.}\ \bibnamefont
  {Scherer}},\ }\href@noop {} {\emph {\bibinfo {title} {Sol-Gel Science}}}\
  (\bibinfo  {publisher} {Academic Press},\ \bibinfo {year} {1990})\BibitemShut
  {NoStop}%
\bibitem [{\citenamefont {Barrett}\ \emph {et~al.}(1951)\citenamefont
  {Barrett}, \citenamefont {Joyner},\ and\ \citenamefont
  {Halenda}}]{BJH_method_1951}%
  \BibitemOpen
  \bibfield  {author} {\bibinfo {author} {\bibfnamefont {E.~P.}\ \bibnamefont
  {Barrett}}, \bibinfo {author} {\bibfnamefont {L.~G.}\ \bibnamefont {Joyner}},
  \ and\ \bibinfo {author} {\bibfnamefont {P.~P.}\ \bibnamefont {Halenda}},\
  }\href@noop {} {\bibfield  {journal} {\bibinfo  {journal} {J. Am. Chem.
  Soc.}\ }\textbf {\bibinfo {volume} {73}},\ \bibinfo {pages} {373} (\bibinfo
  {year} {1951})}\BibitemShut {NoStop}%
\bibitem [{\citenamefont {Shaw}(1987)}]{Shaw1987}%
  \BibitemOpen
  \bibfield  {author} {\bibinfo {author} {\bibfnamefont {T.~M.}\ \bibnamefont
  {Shaw}},\ }\href {http://link.aps.org/doi/10.1103/PhysRevLett.59.1671}
  {\bibfield  {journal} {\bibinfo  {journal} {Phys. Rev. Lett.}\ }\textbf
  {\bibinfo {volume} {59}},\ \bibinfo {pages} {1671} (\bibinfo {year}
  {1987})}\BibitemShut {NoStop}%
\bibitem [{\citenamefont {Prat}(2002)}]{Prat2002}%
  \BibitemOpen
  \bibfield  {author} {\bibinfo {author} {\bibfnamefont {M.}~\bibnamefont
  {Prat}},\ }\href {\doibase http://dx.doi.org/10.1016/S1385-8947(01)00283-2}
  {\bibfield  {journal} {\bibinfo  {journal} {Chemical Engineering Journal}\
  }\textbf {\bibinfo {volume} {86}},\ \bibinfo {pages} {153} (\bibinfo {year}
  {2002})}\BibitemShut {NoStop}%
\bibitem [{\citenamefont {Lehmann}\ \emph {et~al.}(2008)\citenamefont
  {Lehmann}, \citenamefont {Assouline},\ and\ \citenamefont
  {Or}}]{LehmannOr2008}%
  \BibitemOpen
  \bibfield  {author} {\bibinfo {author} {\bibfnamefont {P.}~\bibnamefont
  {Lehmann}}, \bibinfo {author} {\bibfnamefont {S.}~\bibnamefont {Assouline}},
  \ and\ \bibinfo {author} {\bibfnamefont {D.}~\bibnamefont {Or}},\ }\href
  {\doibase 10.1103/PhysRevE.77.056309} {\bibfield  {journal} {\bibinfo
  {journal} {Phys. Rev. E}\ }\textbf {\bibinfo {volume} {77}},\ \bibinfo
  {pages} {056309} (\bibinfo {year} {2008})}\BibitemShut {NoStop}%
\bibitem [{\citenamefont {Wallacher}\ \emph {et~al.}(2004)\citenamefont
  {Wallacher}, \citenamefont {K\"unzner}, \citenamefont {Kovalev},
  \citenamefont {Knorr},\ and\ \citenamefont {Knorr}}]{Wallacher2004}%
  \BibitemOpen
  \bibfield  {author} {\bibinfo {author} {\bibfnamefont {D.}~\bibnamefont
  {Wallacher}}, \bibinfo {author} {\bibfnamefont {N.}~\bibnamefont
  {K\"unzner}}, \bibinfo {author} {\bibfnamefont {D.}~\bibnamefont {Kovalev}},
  \bibinfo {author} {\bibfnamefont {N.}~\bibnamefont {Knorr}}, \ and\ \bibinfo
  {author} {\bibfnamefont {K.}~\bibnamefont {Knorr}},\ }\href {\doibase
  10.1103/PhysRevLett.92.195704} {\bibfield  {journal} {\bibinfo  {journal}
  {Phys. Rev. Lett.}\ }\textbf {\bibinfo {volume} {92}},\ \bibinfo {pages}
  {195704} (\bibinfo {year} {2004})}\BibitemShut {NoStop}%
\bibitem [{\citenamefont {Ravikovitch}\ and\ \citenamefont
  {Neimark}(2002)}]{Ravikovitch2002}%
  \BibitemOpen
  \bibfield  {author} {\bibinfo {author} {\bibfnamefont {P.~I.}\ \bibnamefont
  {Ravikovitch}}\ and\ \bibinfo {author} {\bibfnamefont {A.~V.}\ \bibnamefont
  {Neimark}},\ }\href {http://dx.doi.org/10.1021/la026140z} {\bibfield
  {journal} {\bibinfo  {journal} {Langmuir}\ }\textbf {\bibinfo {volume}
  {18}},\ \bibinfo {pages} {9830} (\bibinfo {year} {2002})}\BibitemShut
  {NoStop}%
\bibitem [{\citenamefont {Schofield}(1948)}]{Schofield1948}%
  \BibitemOpen
  \bibfield  {author} {\bibinfo {author} {\bibfnamefont {R.~K.}\ \bibnamefont
  {Schofield}},\ }\href@noop {} {\bibfield  {journal} {\bibinfo  {journal}
  {Discuss. Faraday Soc.}\ }\textbf {\bibinfo {volume} {3}},\ \bibinfo {pages}
  {105} (\bibinfo {year} {1948})}\BibitemShut {NoStop}%
\bibitem [{\citenamefont {Or}\ and\ \citenamefont {Tuller}(2002)}]{Or2002}%
  \BibitemOpen
  \bibfield  {author} {\bibinfo {author} {\bibfnamefont {D.}~\bibnamefont
  {Or}}\ and\ \bibinfo {author} {\bibfnamefont {M.}~\bibnamefont {Tuller}},\
  }\href {http://dx.doi.org/10.1029/2001WR000282} {\bibfield  {journal}
  {\bibinfo  {journal} {Water Resour. Res.}\ }\textbf {\bibinfo {volume}
  {38}},\ \bibinfo {pages} {19} (\bibinfo {year} {2002})}\BibitemShut {NoStop}%
\bibitem [{\citenamefont {Sarkisov}\ and\ \citenamefont
  {Monson}(2001)}]{Sarkisov2001}%
  \BibitemOpen
  \bibfield  {author} {\bibinfo {author} {\bibfnamefont {L.}~\bibnamefont
  {Sarkisov}}\ and\ \bibinfo {author} {\bibfnamefont {P.~A.}\ \bibnamefont
  {Monson}},\ }\href@noop {} {\bibfield  {journal} {\bibinfo  {journal}
  {Langmuir}\ }\textbf {\bibinfo {volume} {17}},\ \bibinfo {pages} {7600}
  (\bibinfo {year} {2001})}\BibitemShut {NoStop}%
\bibitem [{\citenamefont {Grosman}\ and\ \citenamefont
  {Ortega}(2011)}]{Grosman2011}%
  \BibitemOpen
  \bibfield  {author} {\bibinfo {author} {\bibfnamefont {A.}~\bibnamefont
  {Grosman}}\ and\ \bibinfo {author} {\bibfnamefont {C.}~\bibnamefont
  {Ortega}},\ }\href {http://dx.doi.org/10.1021/la104777y} {\bibfield
  {journal} {\bibinfo  {journal} {Langmuir}\ }\textbf {\bibinfo {volume}
  {27}},\ \bibinfo {pages} {2364} (\bibinfo {year} {2011})}\BibitemShut
  {NoStop}%
\bibitem [{\citenamefont {Bonnet}\ \emph {et~al.}(2013)\citenamefont {Bonnet},
  \citenamefont {Melich}, \citenamefont {Puech},\ and\ \citenamefont
  {Wolf}}]{Bonnet2013}%
  \BibitemOpen
  \bibfield  {author} {\bibinfo {author} {\bibfnamefont {F.}~\bibnamefont
  {Bonnet}}, \bibinfo {author} {\bibfnamefont {M.}~\bibnamefont {Melich}},
  \bibinfo {author} {\bibfnamefont {L.}~\bibnamefont {Puech}}, \ and\ \bibinfo
  {author} {\bibfnamefont {P.~E.}\ \bibnamefont {Wolf}},\ }\href
  {http://stacks.iop.org/0295-5075/101/i=1/a=16010} {\bibfield  {journal}
  {\bibinfo  {journal} {Europhysics Letters}\ }\textbf {\bibinfo {volume}
  {101}},\ \bibinfo {pages} {16010} (\bibinfo {year} {2013})}\BibitemShut
  {NoStop}%
\bibitem [{\citenamefont {Sarkar}\ \emph {et~al.}(1994)\citenamefont {Sarkar},
  \citenamefont {Chaudhuri}, \citenamefont {Wang}, \citenamefont {Kirkbir},\
  and\ \citenamefont {Murata}}]{Sarkar1994}%
  \BibitemOpen
  \bibfield  {author} {\bibinfo {author} {\bibfnamefont {A.}~\bibnamefont
  {Sarkar}}, \bibinfo {author} {\bibfnamefont {S.}~\bibnamefont {Chaudhuri}},
  \bibinfo {author} {\bibfnamefont {S.}~\bibnamefont {Wang}}, \bibinfo {author}
  {\bibfnamefont {F.}~\bibnamefont {Kirkbir}}, \ and\ \bibinfo {author}
  {\bibfnamefont {H.}~\bibnamefont {Murata}},\ }\href
  {http://dx.doi.org/10.1007/BF00486366} {\bibfield  {journal} {\bibinfo
  {journal} {Journal of Sol-Gel Science and Technology}\ }\textbf {\bibinfo
  {volume} {2}},\ \bibinfo {pages} {865} (\bibinfo {year} {1994})}\BibitemShut
  {NoStop}%
\bibitem [{\citenamefont {Scherer}\ and\ \citenamefont
  {Smith}(1995)}]{Scherer1995}%
  \BibitemOpen
  \bibfield  {author} {\bibinfo {author} {\bibfnamefont {G.~W.}\ \bibnamefont
  {Scherer}}\ and\ \bibinfo {author} {\bibfnamefont {D.~M.}\ \bibnamefont
  {Smith}},\ }\href
  {http://www.sciencedirect.com/science/article/B6TXM-3YB50S7-1/2/56f9df8968dbf7e2c8428b25a9b549fc}
  {\bibfield  {journal} {\bibinfo  {journal} {Journal of Non-Crystalline
  Solids}\ }\textbf {\bibinfo {volume} {189}},\ \bibinfo {pages} {197}
  (\bibinfo {year} {1995})}\BibitemShut {NoStop}%
\bibitem [{\citenamefont {Cohan}(1944)}]{Cohan1944}%
  \BibitemOpen
  \bibfield  {author} {\bibinfo {author} {\bibfnamefont {L.~H.}\ \bibnamefont
  {Cohan}},\ }\href {\doibase 10.1021/ja01229a028} {\bibfield  {journal}
  {\bibinfo  {journal} {Journal of the American Chemical Society}\ }\textbf
  {\bibinfo {volume} {66}},\ \bibinfo {pages} {98} (\bibinfo {year}
  {1944})}\BibitemShut {NoStop}%
\bibitem [{\citenamefont {Wheeler}\ and\ \citenamefont
  {Stroock}(2008)}]{Wheeler2008}%
  \BibitemOpen
  \bibfield  {author} {\bibinfo {author} {\bibfnamefont {T.~D.}\ \bibnamefont
  {Wheeler}}\ and\ \bibinfo {author} {\bibfnamefont {A.~D.}\ \bibnamefont
  {Stroock}},\ }\href@noop {} {\bibfield  {journal} {\bibinfo  {journal}
  {Nature}\ }\textbf {\bibinfo {volume} {455}},\ \bibinfo {pages} {208}
  (\bibinfo {year} {2008})}\BibitemShut {NoStop}%
\bibitem [{\citenamefont {Scherer}(1990)}]{Scherer1990}%
  \BibitemOpen
  \bibfield  {author} {\bibinfo {author} {\bibfnamefont {G.~W.}\ \bibnamefont
  {Scherer}},\ }\href {\doibase 10.1111/j.1151-2916.1990.tb05082.x} {\bibfield
  {journal} {\bibinfo  {journal} {Journal of the American Ceramic Society}\
  }\textbf {\bibinfo {volume} {73}},\ \bibinfo {pages} {3} (\bibinfo {year}
  {1990})}\BibitemShut {NoStop}%
\bibitem [{\citenamefont {Debenedetti}(1996)}]{Debenedetti1996}%
  \BibitemOpen
  \bibfield  {author} {\bibinfo {author} {\bibfnamefont {P.~G.}\ \bibnamefont
  {Debenedetti}},\ }\href@noop {} {\emph {\bibinfo {title} {Metastable Liquids:
  Concepts and Principles}}},\ edited by\ \bibinfo {editor} {\bibfnamefont
  {J.~M.}\ \bibnamefont {Prausnitz}}\ and\ \bibinfo {editor} {\bibfnamefont
  {L.}~\bibnamefont {Brewer}}\ (\bibinfo  {publisher} {Princeton University
  Press},\ \bibinfo {year} {1996})\BibitemShut {NoStop}%
\bibitem [{\citenamefont {Bruschi}\ \emph {et~al.}(2010)\citenamefont
  {Bruschi}, \citenamefont {Mistura}, \citenamefont {Liu}, \citenamefont {Lee},
  \citenamefont {G\"{o}sele},\ and\ \citenamefont {Coasne}}]{Bruschi2010}%
  \BibitemOpen
  \bibfield  {author} {\bibinfo {author} {\bibfnamefont {L.}~\bibnamefont
  {Bruschi}}, \bibinfo {author} {\bibfnamefont {G.}~\bibnamefont {Mistura}},
  \bibinfo {author} {\bibfnamefont {L.}~\bibnamefont {Liu}}, \bibinfo {author}
  {\bibfnamefont {W.}~\bibnamefont {Lee}}, \bibinfo {author} {\bibfnamefont
  {U.}~\bibnamefont {G\"{o}sele}}, \ and\ \bibinfo {author} {\bibfnamefont
  {B.}~\bibnamefont {Coasne}},\ }\href {\doibase 10.1021/la1011082} {\bibfield
  {journal} {\bibinfo  {journal} {Langmuir}\ }\textbf {\bibinfo {volume}
  {26}},\ \bibinfo {pages} {11894} (\bibinfo {year} {2010})}\BibitemShut
  {NoStop}%
\bibitem [{Note1()}]{Note1}%
  \BibitemOpen
  \bibinfo {note} {See Supplemental Material}\BibitemShut {NoStop}%
\bibitem [{\citenamefont {Caupin}\ and\ \citenamefont
  {Herbert}(2006)}]{Caupin2006}%
  \BibitemOpen
  \bibfield  {author} {\bibinfo {author} {\bibfnamefont {F.}~\bibnamefont
  {Caupin}}\ and\ \bibinfo {author} {\bibfnamefont {E.}~\bibnamefont
  {Herbert}},\ }\href
  {http://www.sciencedirect.com/science/article/B6X19-4MG6PCC-1/2/5e3a129aca767571cc074275dd4d89bd}
  {\bibfield  {journal} {\bibinfo  {journal} {Comptes Rendus Physique}\
  }\textbf {\bibinfo {volume} {7}},\ \bibinfo {pages} {1000} (\bibinfo {year}
  {2006})}\BibitemShut {NoStop}%
\bibitem [{\citenamefont {Cross}\ and\ \citenamefont
  {Hohenberg}(1993)}]{Cross1993}%
  \BibitemOpen
  \bibfield  {author} {\bibinfo {author} {\bibfnamefont {M.~C.}\ \bibnamefont
  {Cross}}\ and\ \bibinfo {author} {\bibfnamefont {P.~C.}\ \bibnamefont
  {Hohenberg}},\ }\href {http://link.aps.org/doi/10.1103/RevModPhys.65.851}
  {\bibfield  {journal} {\bibinfo  {journal} {Rev. Mod. Phys.}\ }\textbf
  {\bibinfo {volume} {65}},\ \bibinfo {pages} {851} (\bibinfo {year}
  {1993})}\BibitemShut {NoStop}%
\bibitem [{\citenamefont {Jaeger}\ \emph {et~al.}(1989)\citenamefont {Jaeger},
  \citenamefont {Liu},\ and\ \citenamefont {Nagel}}]{Jaeger1989}%
  \BibitemOpen
  \bibfield  {author} {\bibinfo {author} {\bibfnamefont {H.~M.}\ \bibnamefont
  {Jaeger}}, \bibinfo {author} {\bibfnamefont {C.-h.}\ \bibnamefont {Liu}}, \
  and\ \bibinfo {author} {\bibfnamefont {S.~R.}\ \bibnamefont {Nagel}},\ }\href
  {\doibase 10.1103/PhysRevLett.62.40} {\bibfield  {journal} {\bibinfo
  {journal} {Phys. Rev. Lett.}\ }\textbf {\bibinfo {volume} {62}},\ \bibinfo
  {pages} {40} (\bibinfo {year} {1989})}\BibitemShut {NoStop}%
\bibitem [{\citenamefont {Biot}(1941)}]{Biot1941}%
  \BibitemOpen
  \bibfield  {author} {\bibinfo {author} {\bibfnamefont {M.~A.}\ \bibnamefont
  {Biot}},\ }\href {\doibase 10.1063/1.1712886} {\bibfield  {journal} {\bibinfo
   {journal} {Journal of Applied Physics}\ }\textbf {\bibinfo {volume} {12}},\
  \bibinfo {pages} {155} (\bibinfo {year} {1941})}\BibitemShut {NoStop}%
\bibitem [{\citenamefont {Davitt}\ \emph {et~al.}(2010)\citenamefont {Davitt},
  \citenamefont {Arvengas},\ and\ \citenamefont {Caupin}}]{Davitt2010}%
  \BibitemOpen
  \bibfield  {author} {\bibinfo {author} {\bibfnamefont {K.}~\bibnamefont
  {Davitt}}, \bibinfo {author} {\bibfnamefont {A.}~\bibnamefont {Arvengas}}, \
  and\ \bibinfo {author} {\bibfnamefont {F.}~\bibnamefont {Caupin}},\ }\href
  {http://stacks.iop.org/0295-5075/90/i=1/a=16002} {\bibfield  {journal}
  {\bibinfo  {journal} {EPL (Europhysics Letters)}\ }\textbf {\bibinfo {volume}
  {90}},\ \bibinfo {pages} {16002} (\bibinfo {year} {2010})}\BibitemShut
  {NoStop}%
\bibitem [{\citenamefont {Crank}(1979)}]{Crank1979}%
  \BibitemOpen
  \bibfield  {author} {\bibinfo {author} {\bibfnamefont {J.}~\bibnamefont
  {Crank}},\ }\href@noop {} {\emph {\bibinfo {title} {The mathematics of
  diffusion}}}\ (\bibinfo  {publisher} {Oxford University Press},\ \bibinfo
  {year} {1979})\BibitemShut {NoStop}%
\bibitem [{Note2()}]{Note2}%
  \BibitemOpen
  \bibinfo {note} {See Supplemental Movie}\BibitemShut {NoStop}%
\bibitem [{\citenamefont {Stern}(1954)}]{Stern1954}%
  \BibitemOpen
  \bibfield  {author} {\bibinfo {author} {\bibfnamefont {K.~H.}\ \bibnamefont
  {Stern}},\ }\href {\doibase 10.1021/cr60167a003} {\bibfield  {journal}
  {\bibinfo  {journal} {Chemical Reviews}\ }\textbf {\bibinfo {volume} {54}},\
  \bibinfo {pages} {79} (\bibinfo {year} {1954})}\BibitemShut {NoStop}%
\bibitem [{\citenamefont {Prat}\ and\ \citenamefont
  {Bouleux}(1999)}]{Prat1999}%
  \BibitemOpen
  \bibfield  {author} {\bibinfo {author} {\bibfnamefont {M.}~\bibnamefont
  {Prat}}\ and\ \bibinfo {author} {\bibfnamefont {F.}~\bibnamefont {Bouleux}},\
  }\href {http://link.aps.org/doi/10.1103/PhysRevE.60.5647} {\bibfield
  {journal} {\bibinfo  {journal} {Phys. Rev. E}\ }\textbf {\bibinfo {volume}
  {60}},\ \bibinfo {pages} {5647} (\bibinfo {year} {1999})}\BibitemShut
  {NoStop}%
\bibitem [{\citenamefont {Huppert}\ and\ \citenamefont
  {Neufeld}(2014)}]{Huppert2014}%
  \BibitemOpen
  \bibfield  {author} {\bibinfo {author} {\bibfnamefont {H.~E.}\ \bibnamefont
  {Huppert}}\ and\ \bibinfo {author} {\bibfnamefont {J.~A.}\ \bibnamefont
  {Neufeld}},\ }\href {\doibase 10.1146/annurev-fluid-011212-140627} {\bibfield
   {journal} {\bibinfo  {journal} {Annual Review of Fluid Mechanics}\ }\textbf
  {\bibinfo {volume} {46}},\ \bibinfo {pages} {255} (\bibinfo {year}
  {2014})}\BibitemShut {NoStop}%
\bibitem [{\citenamefont {Dixon}\ \emph {et~al.}(1984)\citenamefont {Dixon},
  \citenamefont {Grace},\ and\ \citenamefont {Tyree}}]{Dixon1984}%
  \BibitemOpen
  \bibfield  {author} {\bibinfo {author} {\bibfnamefont {M.~A.}\ \bibnamefont
  {Dixon}}, \bibinfo {author} {\bibfnamefont {J.}~\bibnamefont {Grace}}, \ and\
  \bibinfo {author} {\bibfnamefont {M.~T.}\ \bibnamefont {Tyree}},\ }\href
  {\doibase 10.1111/1365-3040.ep11592146} {\bibfield  {journal} {\bibinfo
  {journal} {Plant, Cell \& Environment}\ }\textbf {\bibinfo {volume} {7}},\
  \bibinfo {pages} {615} (\bibinfo {year} {1984})}\BibitemShut {NoStop}%
\end{thebibliography}%

\end{document}